\documentclass[pra, aps, twocolumn, groupedaddress, showpacs, superscriptaddress]{revtex4}

\usepackage{slashed}
\usepackage{graphicx}
\usepackage{subfigure}
\usepackage[usenames, dvipsnames]{color}
\usepackage{graphics}
\usepackage{hyperref}
\usepackage{bm}
\usepackage{amsfonts}
\hypersetup{backref,
colorlinks=true,
urlcolor=blue,
linkcolor=blue,
linktoc=page,
citecolor=blue}

\everymath{\displaystyle} 
\begin{document}

\title{Quantum dark solitons as qubits in Bose-Einstein condensates}

\author{Muzzamal I. Shaukat}
\affiliation{CeFEMA, Instituto Superior T\'ecnico, Universidade de Lisboa, Lisboa, Portugal}
\affiliation{University of Engineering and Technology, Lahore (RCET Campus), Pakistan}
\email{muzzamalshaukat@gmail.com}

\author{Eduardo. V. Castro}
\affiliation{CeFEMA, Instituto Superior T\'ecnico, Universidade de Lisboa, Lisboa, Portugal}

\author{Hugo Ter\c{c}as}

\affiliation{Instituto de Plasmas e Fus\~ao Nuclear, Lisboa, Portugal}
\affiliation{Instituto de Telecomunica\c c\~oes, Lisboa, Portugal}
\email{hugo.tercas@tecnico.ulisboa.pt}


\begin{abstract}
We study the possibility of using dark-solitons in quasi one dimensional
Bose-Einstein condensates to produce two-level systems (qubits) by exploiting the intrinsic nonlinear and the coherent nature of the matter waves. We
calculate the soliton spectrum and the conditions for a qubit to exist. We also compute the coupling between the phonons and the solitons and investigate the emission rate of the qubit in that case. Remarkably, the qubit lifetime is estimated to be of the order of a few seconds, being only limited by the dark-soliton ``death" due to quantum evaporation.
\end{abstract}

\maketitle

\section{Introduction}

Quantum effects strive to disappear for macroscopic objects. Typically, quantum effects become defamed out into their classical averages, and therefore manipulation of quantum states, relevant for quantum computation, becomes unsustainable at the macroscopic scale. However, Bose-Einstein condensates (BECs) constitute one important exception where quantum effects are perceptible on a macroscopic level. With the advent and rapid developing of laser cooling and trapping of neutral atoms over the past decade, micron-sized atomic gases at ultralow temperatures are routinely formed in the laboratory \cite{henderson2009, donley2001, leanhardt2003}. Moreover, quantum optic techiques allow for an unprecedent versatile and precise control on internal degrees of freedom, putting cold atoms as one of the most prominent candidates to test complex aspects of strongly correlated matter and to applications in quantum information processing \cite{zoller98, greiner2002, orth2008, kalas2008, santamore2008, solenov2008a, solenov2008b, kalas2010}. \par

Quantum information has been introduced in cold atom systems at various levels \cite{porto2003, lundblad2009}. One way consists in defining a qubit (a two-state system) via two internal states of an atom. This approach, however, requires each atom to be addressed separately. A similar problem appears when the qubit is introduced via a set of spatially localized states (e.g. in adjacent wells of an optical lattice potential) of an atom or a BEC. The complication is due to the fact that the number of atoms in a BEC experiment significantly fluctuates from run to run. As a result, any qubit system dependent on the number of atoms becomes problematic. A second way of producing qubits in these systems relies on collective properties of ultracold atoms. Here, a two-state system can be formed by isolating a pair of macroscopic states that are set sufficiently far away from the multiparticle spectrum. At the same time, however, the energy gap between these lowest states must remain small enough to allow measurable dynamics \cite{leggett89}. Experiments performed in the double-well potential configuration are a pioneer example of such approach \cite{dw_2008, andrews}. Nevertheless, despite the appealing similarity with single-particle two-level states, double-well potentials lack to achieve a macroscopic superposition allowing for a measurable dynamics \cite{andrews}. To overcome the superposition issue, a more recent proposal based on BEC superfluid current states in the ring geometry, analogous to the superconducting flux qubit \cite{tinkham2004}, has been discussed \cite{solenov2011}. More recently, the concept of {\it phononic} reservoir via the manipulation of the phononic degrees of freedom has pushed quantum information realizations to another level, comprising the dynamics of impurities immersed in BECs \cite{klein2007, cirone2009, haikka2011, mulansky2011, peotta2013} and reservoir engineering to produce multipartite dark states \cite{stanigel2012, ramos2014, petersen2014, mitsch2014, sollner2014, young2014}. Another important difference in respect to quantum optical system is the possibility to use phononic reservoirs to test non-Markovian effects in many-body systems \cite{ramos2016, vermersch2016}. The implementation of quantum
gates has been recently proposed in Ref. \cite{luiz}.  \par

\begin{figure}[t!]
\includegraphics[scale=0.4]{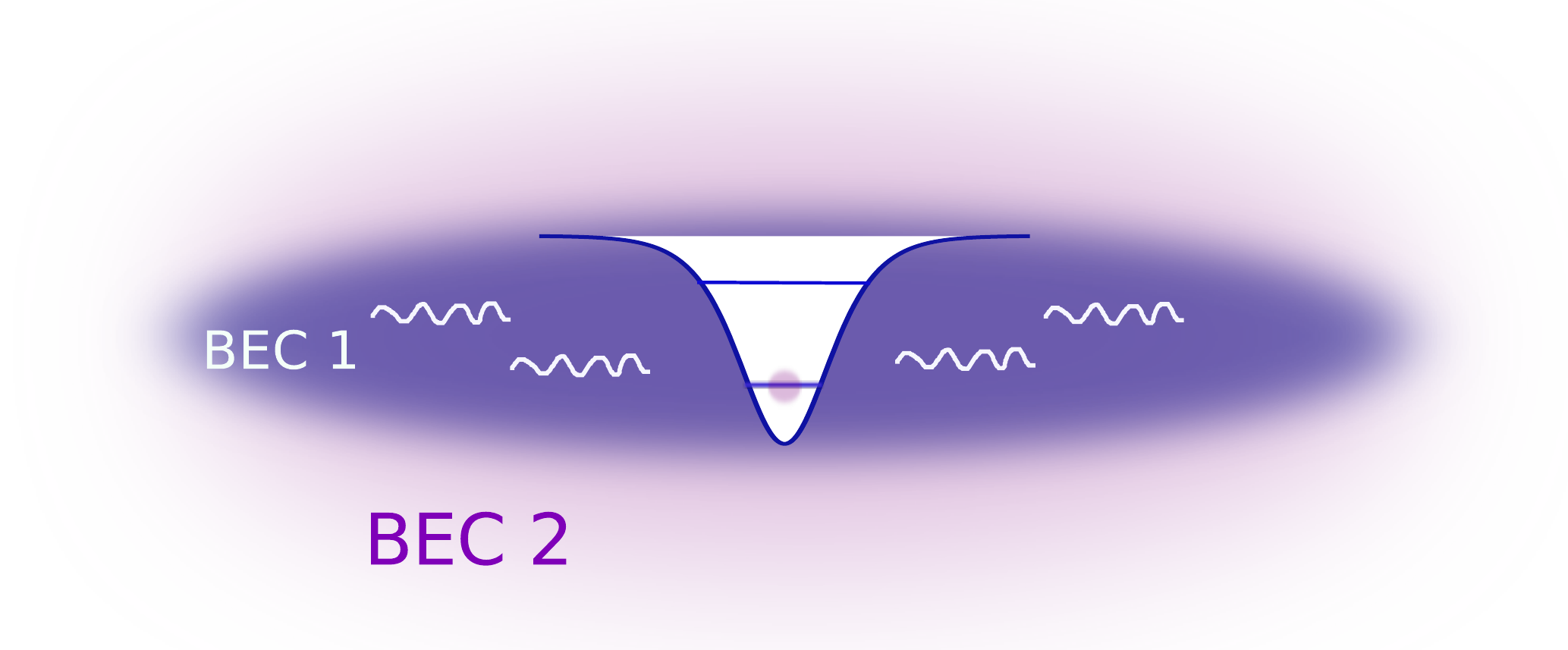}
\caption{(color online) Schematic representation of the problem with two coupled BECs. BEC 1 contains a dark soliton, which acts as a potential to the particles of BEC 2. Under certain circumstances, exactly two bound states can be formed. Due to quantum fluctuations, BEC 1 also support phonons (wiggly lines), which will interact with the dark soliton and, consequently, provide some dephasing.}
\label{fig_scheme}
\end{figure}

Another important family of macroscopic structures in BECs with potential applications in quantum information are the so-called dark-solitons (DS). They consist of nonlinear localized depressions in a quasi-1D BEC that emerge due to a precise balance between the dispersive and nonlinear effects in the system \cite{kivshar, burger, Denschlag, anderson}, being also ubiquitous in nonlinear optics \cite{Krakel}, shallow liquids \cite{Denardo}, magnetic films \cite{Chen}. Quasi one-dimensional BECs with repulsive interatomic interaction are prone to inception of dark solitons by various methods, like imprinting spatial phase distribution  \cite{Denschlag}, inducing density defects in BEC \cite{Dutton}, and by collision of two condensates \cite{Reinhardt, Scott}.
The stability and dynamics of DS in BECs have been a subject of intense research over the last decade \cite{jackson, dziarmaga}. Recent activity in the field involve studies on the collective aspects of the so-called {\it soliton gases} \cite{gael2005}, putting dark solitons as a good candidate to investigate many-body physics \cite{tercas}. \par

In this paper, we combine the intrinsic nonlinearity in quasi one-dimensional BECs to construct two-level states (qubits) with dark solitons. As we will show, thanks to the unique properties of the DS spectrum, perfectly isolated two-level states are possible to construct. As a result, a matter-wave qubit of a few kHz energy gap is achieved. The effect of decoherece due to the presence of phonons (quantum fluctuations around the bacground density) play the role of a proper quantum reservoir. Remarkably, due to their instrisic slow-time dynamics, BEC phonons provide small decoherence rates of few Hz, meaning that under typical experimental conditions, DS-qubits have a lifetime comparable to the lifetime of the BEC, being only limited by the soliton quantum diffusion (``evaporation"). As we show below, this effect is not critical and the qubit is still robust within the 100 ms$-$0.1 s time scale.


The paper is organized as follows: In sec. II, the properties of a single DS in a quasi-1D BEC immersed in a second condensate are derived. We start with the set of coupled Gross Pitaevskii equations and find under which conditions DSs can uniquevocally define a two level atom (qubit). In Sec. III, we compute the coupling between phonons and DSs. Sec. IV discusses the Weisskopf-Wigner theory to determine the emission rate of the qubit. Some discussion and conclusions about the implications of our proposal in practical quantum information protocols are stated in Sec. V. 

\section{Mean-field equations and the dark-soliton qubit}

\begin{figure}[tbp]
\includegraphics[scale=0.5]{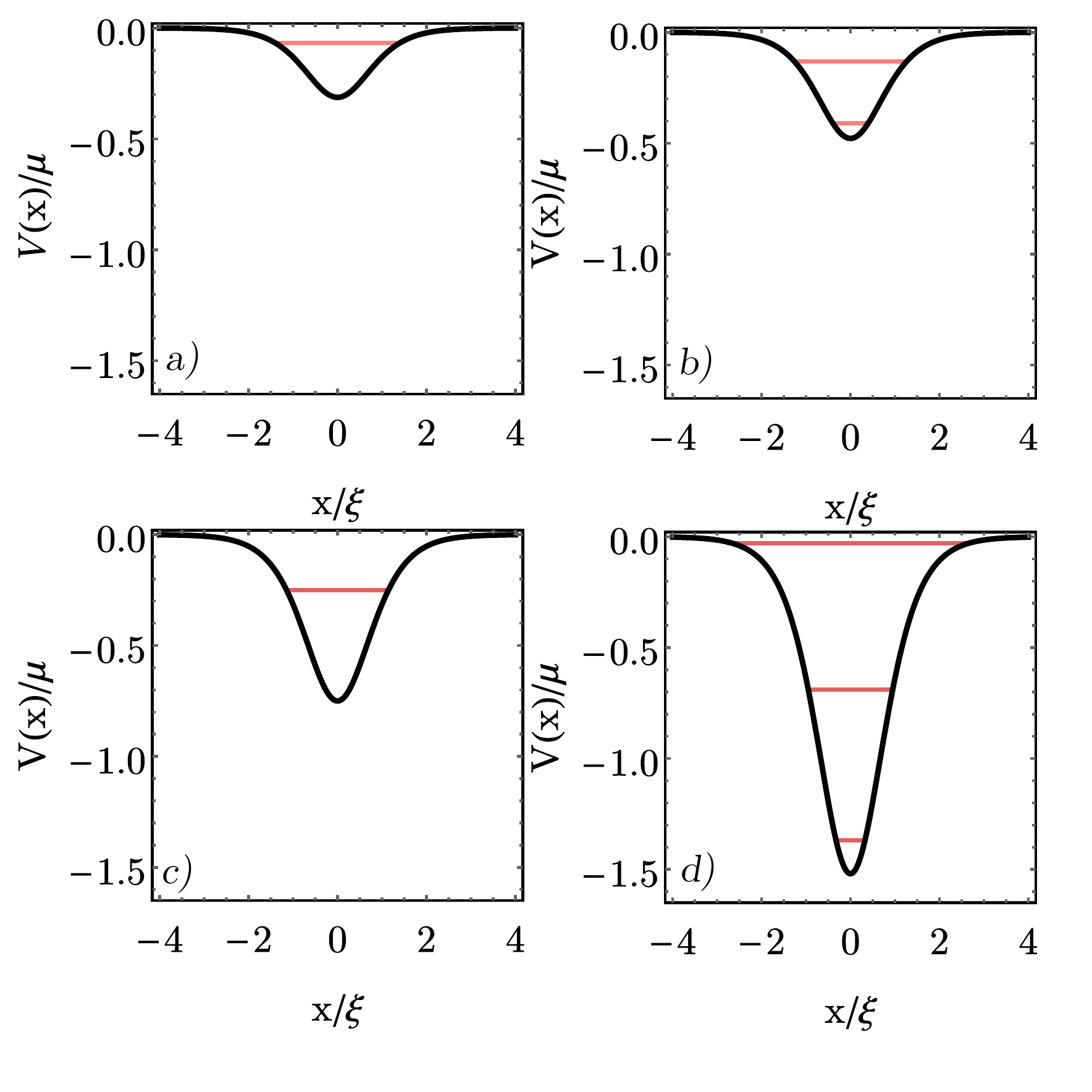}
\caption{(color online) Dark soliton acting as a binding potential for the free particles. Panel a) depicts a single bound state obtained for $\nu < 1/3$, while panel b) illustrates the case of a two-level system (qubit) case obtaind for $1/3\leq\nu < 4/5$. Panel c) illustrates the patological case $\nu=1/2$ for which a degeneracy in the two-level system is obtained. A qutrit case is depicted in panel d) for $\nu=0.83$.}
\label{fig_spectrum}
\end{figure}

We consider two-coupled quasi-1D BECs. A quasi 1D gas is produced when the transverse dimension of the trap is larger than or of the order of the $s$-wave scattering length and, at the same time, much smaller than the longitudinal extension \cite{perez, carr}. At the mean field level, the dynamics of the system is thus governed by the time-dependent coupled Gross Pitaevskii equations 
\begin{equation}
i\hbar \frac{\partial \psi _{1}}{\partial t}=-\frac{\hbar ^{2}}{2m}\frac{
\partial^{2} \psi _{2}}{\partial x^{2}}+g_{11}\left\vert \psi _{1}\right\vert ^{2}\psi
_{1}+g_{12}\left\vert \psi _{2}\right\vert ^{2}\psi _{1}  \label{gp1.}
\end{equation}%
\begin{equation}
i\hbar \frac{\partial \psi _{2}}{\partial t}=-\frac{\hbar ^{2}}{2m}\frac{
\partial^{2} \psi _{2}}{\partial x^{2}}+g_{22}\left\vert \psi _{2}\right\vert ^{2}\psi
_{2}+g_{21}\left\vert \psi _{1}\right\vert ^{2}\psi _{2}  \label{gp2}
\end{equation}%
where $g_{11}$ ($g_{22}$) is the one-dimensional coupling strength between particles in BEC$_1$ (BEC$_2$) and $g_{12}=g_{21}$ is the inter-particle coupling constant, $\hbar $ is the Planck constant, $m$ is the mass of the atomic species. We restrict the discussion to repulsive interatomic interacions, $g_{11}(g_{22})>0$. In what follow, we assume that a dark soliton is present in BEC$_1$ and $%
g_{22}\ll g_{12}\leq g_{11}$ such that particles in BEC$_2$ do not interact and can
therefore be regarded as a set of free interacting particles (see Fig. \ref{fig_scheme}). Thus, Eq. (\ref{gp2}) can be written as
\begin{equation}
i\hbar \frac{\partial \psi _{2}}{\partial t}=-\frac{\hbar ^{2}}{2m}\frac{%
\partial^{2} \psi _{2}}{\partial x^{2}}+g_{21}\left\vert \psi _{\rm sol}\right\vert
^{2}\psi _{2},  \label{sch. eq. without soliton}
\end{equation}
where the soliton profile, resulting a singular nonlinear solution to Eq. (\ref{gp1.}), is given by \cite{Muryshev, huang, zakharov72, zakharov73}
\begin{equation}
\psi _{\rm sol}(x)=\sqrt{n_{0}}\tanh \left( \frac{x}{\xi }\right). 
\label{eq_sol}
\end{equation}
Here, $n_{0}$ is the background density which is typically of the order of $10^{8}$ m$^{-1}$ in elongated BECs,
and the healing length $\xi =\hbar /\sqrt{mn_{0}g_{11}}$ is of the order $(0.2-0.7)$ $\mu$m. We also consider the experimentally accesible trap frequencies $\omega_{r}=2\pi \times (1-5)$ kHz $\gg \omega _{z}=2\pi \times (15-730)$ Hz and the
corresponding length amount to be the value $l_{z}=(0.6-3.9)$ $\mu$m \cite{parker}. Notice that the previous results can be easily generalized for the case of a gray solitons (i.e solitons traveling with speed $v$) by replacing Eq. (\ref{eq_sol}) by 
\begin{equation}
\psi _{\rm sol}(x)=\sqrt{n_{0}}\left[i\theta+\frac{1}{\gamma}\tanh \left( \frac{x}{\xi\gamma}\right)\right], 
\label{eq_gray sol}
\end{equation}
where $\theta=v/c_s$, $\gamma=(1-\theta^2)^{-1/2}$, and $c_s=\sqrt{gn_0/m}$ is the BEC sound speed \cite{tercas, pelinovsky, wadkin}. Therefore, the time-independent version of Eq. (\ref{sch. eq. without soliton}) reads 
\begin{equation}
E'\psi _{2}=-\frac{\hbar ^{2}}{2m}\frac{%
\partial^{2} \psi _{2}}{\partial x^{2}}-g_{21}n_{0}{\rm sech}^{2}\left( \frac{x}{\xi }\right) \psi _{2},
\label{eq_reflectionless1}
\end{equation}%
where $E'=E-g_{21}n_{0}$. Here, the dark soliton act as a potential for the particles of the reservoir. Analytical solutions to Eq. (\ref{eq_reflectionless1}) can be obtained by casting the  potential term in the form of \cite{john}
\begin{equation}
V(x)=-\frac{\hbar ^{2}}{2m\xi ^{2}}\nu (\nu +1){\rm sech}^{2}\left( 
\frac{x}{\xi }\right),  
\label{Potent.}
\end{equation}
where $\nu=\left(-1+\sqrt{1+4g_{12}/g_{11}}\right)/2$. The particular case of $\nu$ being a positive integer corresponds to the important case of the {\it reflectionless} potential \cite{john}, for which an incident wave is totally transmitted. For the more general case considered here, the energy spectrum associated to the potential in Eq. (\ref{Potent.}) reads
\begin{equation}
E_{n}^{^{\prime }}=-\frac{\hbar ^{2}}{2m\xi ^{2}}(\nu -n)^{2},
\label{energy eigen states}
\end{equation}
where $n$ is an integer. The number of bound states is given by $n_{\rm bound}=\lfloor 1+\nu+\sqrt{\nu(\nu+1)}\rfloor$, where $\lfloor \cdot\rfloor$ denotes the integer part. A two-level system (qubit) can be perfectly isolated when the value of $\nu $ ranges as
\[
\frac{1}{3}\leq \nu <\frac{4}{5}. 
\]%
At the critical point $\nu =1/2$ the two-energy levels merge and the qubit is ill-defined. Finally, for $\nu \geq 4/5$, three-level systems (qutrits) can also be formed, but this case is out of the scope of the present work and will be discussed in a separate publication. The features of the spectrum (\ref{energy eigen states}) are illustrated in Fig. \ref{fig_spectrum}.
\section{Qubit-Phonon interaction}

The mean-field soliton soluton in Eq. (\ref{eq_sol}) is accompained by quantum fluctuations (phonons). In that case, the total wavefunction is given by $\psi_1(x)=\psi_{\rm sol}(x)+\delta \psi_1(x)$, where
\begin{equation}
\delta \psi_1(x)=\sum_k \left(u_k(x) b_k e^{ikx} +v_k(x)^*b_k^\dagger e^{-ikx} \right),
\label{eq_dec1}
\end{equation}
with $b_k$ denoting bosonic operators satisfying the commutation relation $[b_k,b_q^\dagger]=\delta_{kq}$. $u_k(x)$ and $v_k(x)$ are amplitudes veryfing the normalization condition $\vert u_k(x)\vert ^2 -\vert v_k(x)\vert ^2=1$ and are explicitly given by
\cite{Dziarmaga04}  
\begin{eqnarray*}
&&\left. u_{k}(x)=\sqrt{\frac{1}{4\pi \xi }}\frac{\mu }{\epsilon _{k}}%
\right. \times \\
&&\left. \left[ \left( (k\xi )^{2}+\frac{2\epsilon _{k}}{\mu }\right) \left( 
\frac{k\xi }{2}+i\tanh \left( \frac{x}{\xi }\right) \right) +\frac{k\xi }{%
\cosh ^{2}\left( \frac{x}{\xi }\right) }\right] \right. ,
\end{eqnarray*}%
and 
\begin{eqnarray*}
&&\left. v_{k}(x)=\sqrt{\frac{1}{4\pi \xi }}\frac{\mu }{\epsilon _{k}}%
\right. \times \\
&&\left. \left[ \left( (k\xi )^{2}-\frac{2\epsilon _{k}}{\mu }\right) \left( 
\frac{k\xi }{2}+i\tanh \left( \frac{x}{\xi }\right) \right) +\frac{k\xi }{%
\cosh ^{2}\left( \frac{x}{\xi }\right) }\right] \right. .
\end{eqnarray*}%
Similarly, the particles in BEC$_2$ are eigenstates of Eq. (\ref{eq_reflectionless1}), being therefore spannable in terms of the bosonic operators $a_\ell$ as
\begin{equation}
\psi_2(x)=\sum_{\ell=0}^1 \varphi_\ell(x) a_\ell,
\label{eq_dec2}
\end{equation}
where $\varphi_0(x)={\rm sech}(x/\xi)/(\sqrt{2\xi})$ and $\phi_1(x)=i\sqrt{3}\tanh(x/\xi)\varphi_0(x)$.
The total Hamiltonian may then be written as
\begin{equation}
H=H_{\rm qubit}+H_{\rm ph}+H_{\rm int}. 
\end{equation}
The first term $H_{\rm qubit}$ represents the dark-soliton (qubit) Hamiltonian 
\begin{equation}
H_{\rm qubit}=\hbar \omega _{0}\sigma _{z}  
\label{system Hamilt.}
\end{equation}
where $\omega _{0}=\hbar(2\nu -1)/(2m\xi ^{2})$ is the qubit gap frequency and $\sigma_z=a_1^\dagger a_1- a_0^\dagger a_0$ is the corresponding spin operator. The second term describes the phonon (reservoir) Hamiltonian
\begin{equation}
H_{\rm }=\sum_k \epsilon _{k}b _{k}^{\dagger }b _{k},
\label{envir. hamilt.}
\end{equation}
where the Bogoliubov spectrum is given by $\epsilon _{k}=\mu \xi \sqrt{k^{2}(\xi^{2}k^{2}+2)}$ and $\mu =gn_{0}$ denotes the chemical potential. \par
\begin{figure}
\centering
\includegraphics[scale=0.5]{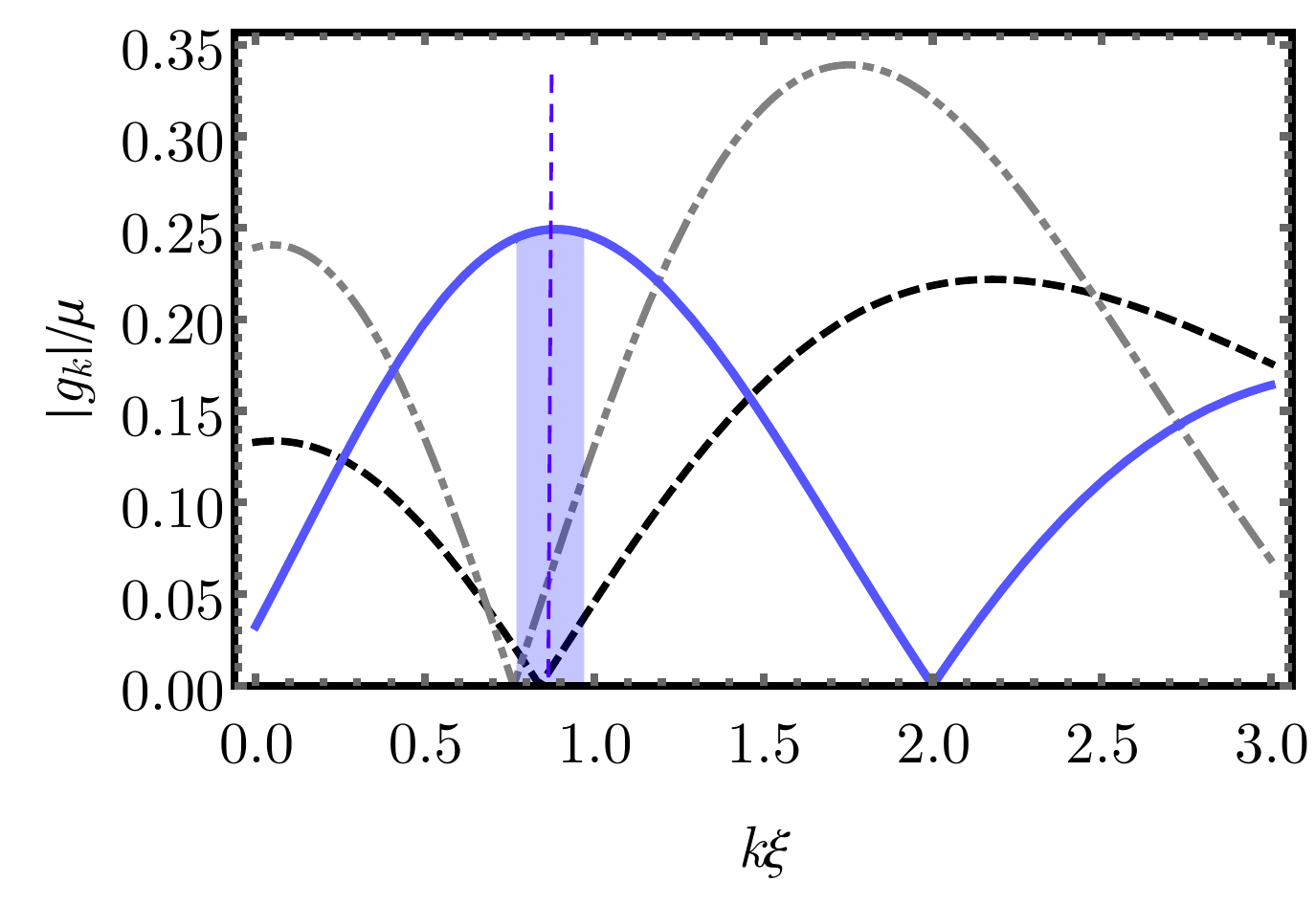}
\caption{(color online) Intra-band $g_{00}(k)$ (dashed line) and $g_{11}(k)$ (dotted-dashed line) and inter-band $g_{10}(k)$ (solid line) coupling functions. Around the resonant values $k\sim \xi^{-1}$ (shadowed region), the interband term is dominant, allowing us to neglect the intra-band terms within the rotating-wave approximation.}
\label{fig_RWA}
\end{figure}

The interaction Hamiltonian $H_{\rm int}$ between qubit and the reservoir is defined as
\begin{equation}
H_{\rm int}=g_{12}\int dx\psi _{2}^{\dag }\psi _{1}^{\dag }\psi _{1}\psi _{2}
\label{Int. Ham.}
\end{equation}
which, with the prescriptions in Eqs. (\ref{eq_dec1}) and (\ref{eq_dec2}), can be decomposed as
\begin{equation}
H_{\rm int}=H_{\rm int}^{(0)}+H_{\rm int}^{(1)}+H_{\rm int}^{(2)}, 
\label{Inter. hamilt.}
\end{equation}
respectivelty containing zero, first and second order terms in the operators $b_k$ and $b_k^\dagger$. Owing to the small depletion of the condensate, and consistent with the Bogoliubov approximation performed in Eq. (\ref{eq_dec1}), we ignore the higher-order term $H_{\rm int}^{(2)}\sim \mathcal{O}(b_{k}^{2})$. The first part of Eq. (\ref{Inter. hamilt.}) corresponds to a Stark shift term of the type
\begin{equation}
H_{\rm int}^{(0)}=g_{12}n_0\delta _{\ell \ell'}a_{\ell}^{\dagger}a_{\ell'}f_{\ell \ell'},
\end{equation}
where $f_{\ell \ell'}=\int dx~\varphi _{\ell}^{\dag }(x)\varphi _{\ell'}(x)\tanh ^{2}\left( 
x/\xi \right) $. The latter can be omitted by renormalizing the qubit frequency as $\widetilde{\omega }_{0}=\omega
_{0}+n_{0}g_{12}.$ In its turn, the first-order term $\mathcal{O}(b_{k})$ is given by
\begin{equation}
H_{\rm int}^{(1)}=\sum_{k}\sum_{\ell,\ell'}a_{\ell}^{\dag }a_{\ell'}\left(
b_{k}g_{\ell,\ell'}(k)+b_{k}^{\dag }g_{\ell,\ell'}(k)^*\right) +{\rm h.c.}
\label{one hamiltonian}
\end{equation}
where 
\begin{equation}
g_{\ell,\ell'}(k) =\sqrt{n_{0}}g_{12}\int dx\varphi _{\ell}^{\dag
}(x)\varphi _{\ell'}(x)\tanh \left( \frac{x}{\xi }\right)e^{ikx} u_{k}
\end{equation}
As we can observe, Eq. (\ref{one hamiltonian}) contains intraband ($\ell=\ell'$) and
interband ($\ell\neq \ell'$) terms. However, for small values of the coupling between the system and the environment, the qubit transition can only be driven by near-resonant phonons, for which te interband coupling amplitude $\vert g_{01}(k)\vert=\vert g_{10}(k)^*\vert$ is much larger that the interband terms $\vert g_{00}(k) \vert$ and $\vert g_{11}(k)\vert$ (see Fig. \ref{fig_RWA}). As such, within the rotating-wave approximation (RWA), we can safely drop the intraband terms to obain 
\begin{equation}
H_{\rm int}^{(1)}=\sum_{k}g(k)\sigma_{+}b_{k}+\sum_{k}g(k)^*\sigma_{-}b_{k}^{\dagger} + {\rm h.c.},
\label{eq_ham_final}
\end{equation}%
where $\sigma_+=a_1^\dagger a_0$, $\sigma_-=a_0^\dagger a_1$ and the coupling constant $g_k\equiv g_{0,1}(k)=-g_{1,0}(k)$ is explicitly given by
\begin{eqnarray*}
g_k &=&\frac{ig_{12}k^{2}\xi ^{3/2}}{80\epsilon _{k}}\sqrt{\frac{%
n_{0}\pi }{6}}(2\mu +8k^{2}\mu \xi ^{2}+15\epsilon _{k}) \\
&&\left( -4+k^{2}\xi ^{2}\right) {\rm csch} \left( \frac{k\pi \xi }{2}\right).
\end{eqnarray*}
We notice that the implementation of the RWA approximation also implied the dropping of the counter-rotating terms proportional to $b_k \sigma_-$ and $b_k^\dagger\sigma_+$ that do not conserve the total number of excitations. The accuracy of such an approximation can be verified {\it a posteriori}, provided that the emission rate $\Gamma$ is much smaller than the qubit transition frequency $\omega_0$. 
\section{Spontaneous decay of the dark-soliton qubit}

\begin{figure}[t!]
\flushleft
\includegraphics[scale=0.41]{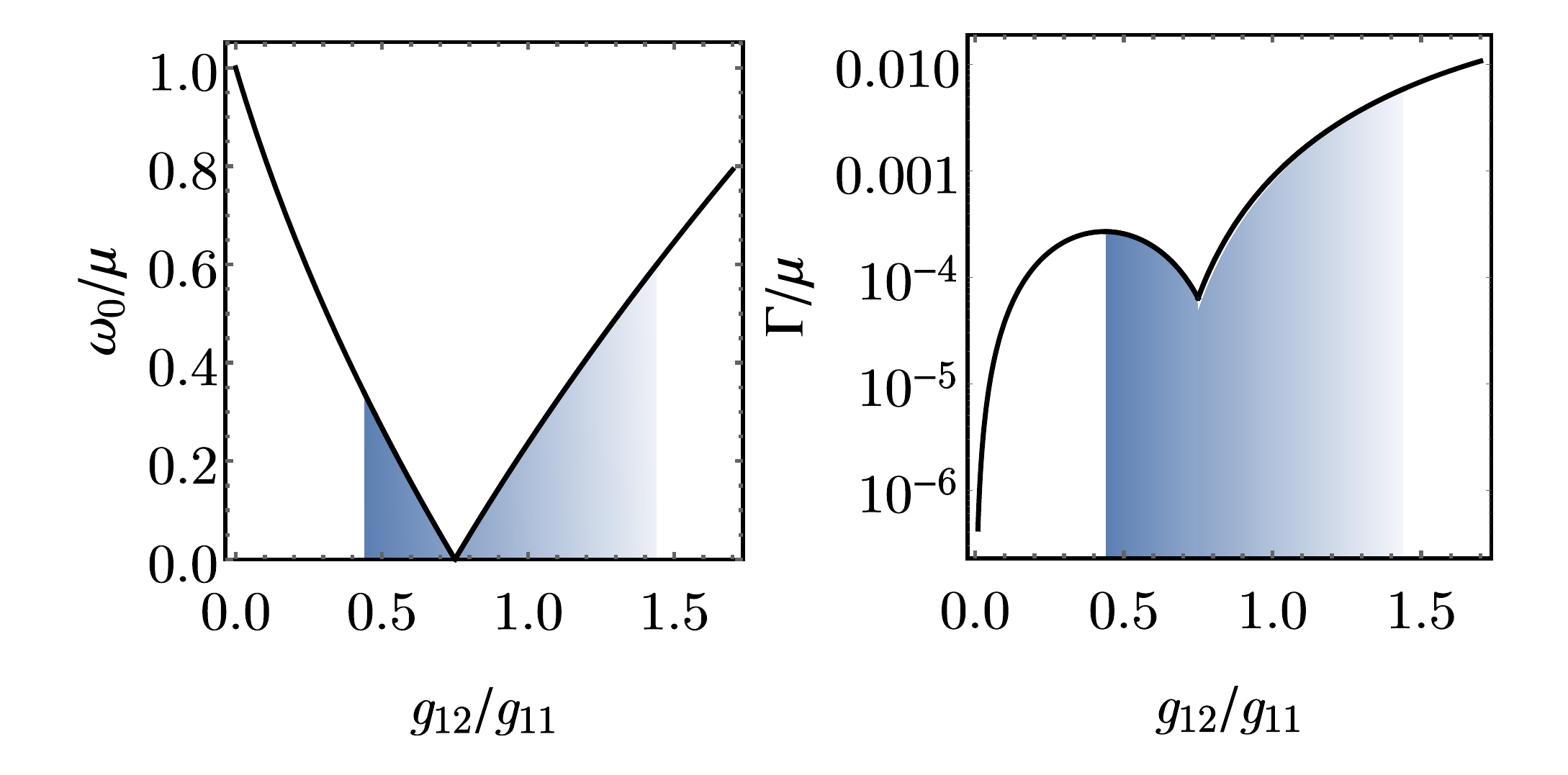}
\caption{(color online) Dependence of the transition frequency $\omega_0$ (panel a)) and decay rate $\Gamma$ (panel b)) on the coupling constant $g_{12}$. The shadowed region corresponds to the range $1/3\leq\nu< 4/5$ for which the quibit can be exactly defined. The case $\nu=1/2$ produces a degenerate two-level system.}
\label{fig_gamma}
\end{figure}

Neglecting the effect of temperature and other external perturbations, the only source of decoherence of a dark-soliton qubit are te sourrounding phonons. Because cold atom experiments are typically very clean, and considering that the zero-temperature approximation is an excellent approximation for quasi-1D BECs \cite{pethick, pitaevskii}, we employ the Wigner-Weisskopf theory in order to compute the lifetime of the qubit. We assume the qubit to be initially in its excited state and the field to be in the vacuum state. Under such conditons, the total system+reservoir wavefunction can be parametrized as 
\begin{equation}
\left\vert \psi (t)\right\rangle =\alpha(t)e^{-i\omega _{0}t}\left\vert
e,0\right\rangle +\sum\limits_{k}\beta_{k}(t)e^{-i\omega _{k}t}\left\vert
g,1_{k}\right\rangle  \label{time depend. eigen fun.}
\end{equation}%
where $\alpha(t)$ and $\beta_k(t)$ are the probability amplitudes. The Wigner-Weisskopf ansatz (\ref{time depend. eigen fun.}) is then let to evolve under the total Hamiltonian in Eq. (\ref{eq_ham_final}), for which the Schr\"odinger equation yields the following evolution of the coefficients
\begin{eqnarray}
\dot \alpha(t) &=&\frac{i}{\hbar }\sum%
\limits_{k}g_{k}e^{-i(\omega _{k}-\omega _{0})t}\beta_{k}(t)
\label{coefficient a} \\
\beta_{k}(t) &=&\frac{i}{\hbar }g_{k}^{\ast }\int_{0}^{t}\alpha(t^{^{\prime
}})e^{i(\omega _{k}-\omega _{0})t^{^{\prime }}}dt^{^{\prime }}.
\label{coefficient b}
\end{eqnarray}
Due to separation of time scales between the phonons and the decay process, we may assume that the coefficient $\alpha(t)$ evolves much slower than $\beta_k(t)$, which allows us to evoque the Born approximation to write
\[
\int_{0}^{t}\alpha(t^{^{\prime }})e^{-i(\omega _{k}-\omega _{0})(t-t^{^{\prime
}})}dt^{^{\prime }}\simeq \alpha(t)\int_{0}^{t}e^{-i(\omega _{k}-\omega
_{0})\tau }d\tau , 
\]%
where $\tau =t-t^{^{\prime }}.$ Moreover, since we expect $\alpha(t)$ to varie at a rate $\Gamma \ll
\omega _{0}$, the relevant decay dynamics is expected to take place at times $t\gg \frac{1}{\omega _{0}}$, which allows us to take upper limit of above integral to $\infty$ (Markov approximation). Therefore, we have %

\begin{equation}
\begin{array}{ccc}
\alpha(t)\int_{0}^{\infty }e^{-i(\omega _{k}-\omega _{0})\tau }d\tau &=&\alpha(t)\pi
\delta (\omega _{k}-\omega _{0})\\
&-&i \alpha(t)\wp \left(\frac{1}{\omega _{k}-\omega _{0}}\right)
\end{array}
\label{time integral}
\end{equation}
where $\wp $ represents the Cauchy principal part describing an additional energy (Lamb) shift. Because it represents a small correction to the qubit energy $\omega_0$, we do not compute its contribution explicitly. Therefore, the excited state amplitude
decays exponentially as 
\begin{equation}
\alpha(t)=e^{-\Gamma t/2}.  \label{Coefficient a + gamma}
\end{equation}
where $\Gamma $ is the population decay rate given as
\begin{eqnarray}
\Gamma &=&\frac{L}{\sqrt{2}\hbar \xi}\int ~ d\omega_k  \frac{\sqrt{1+ \eta_k}}{\eta_k}\vert g_k\vert^2 \delta(\omega_k-\omega_0)\\
&=&\frac{\pi N_{0}g_{12}^{2}}{76800\hbar \mu ^{5}\xi ^{2}\eta_0 \sqrt{%
\frac{\mu +\eta_0 }{\mu }}}\left( -\mu +\eta_0 \right) \left( -5\mu +\eta_0
\right) ^{2}\nonumber \\
&\times&\left( 8\eta_0 +3\mu \left(-2+5\xi \sqrt{\frac{\hbar ^{2}\omega _{0}^{2}}{%
\mu ^{2}\xi ^{2}}}\right) \right) ^{2}\\
&\times&{\rm csch}^{2}\left( \frac{\pi \sqrt{-\mu
+\eta_0 }}{2\sqrt{\mu }}\right)\nonumber
\label{eq_gamma}
\end{eqnarray}
where $\eta_0,_k =\sqrt{\mu ^{2}+\hbar ^{2}\omega _{0,k}^{2}}$. As depicted in Fig. \ref{fig_gamma}, the decay rate $\Gamma$ is orders of magnitude smaller than the qubit gap $\omega_0$, confirming that the evoquing both the RWA and the Born-Markov approximation can also be used for phononic systems. Remarkably, for a quasi-1D of chemical potential of few kHz, we can obtain a qubit lifetime $\tau_{\rm qubit}\sim 1/\Gamma$ of the order of a second, a time comparable to lifetime of the BEC itself. Notice that the value of $g_{12}$ (and consequently the qubit natural frequency $\omega_0$ and lifetime $\tau_{\rm qubit}$) can be experimentally tunned with the help of Feshbach resonances. The only immediate limitation to the performance of our proposal may be related to the dark-soliton quantum diffusion \cite{Dziarmaga04}. Since they interact with the background phonons, they are expected to evaporate within the time scale $\tau_{\rm diffusion}=8\xi/c_s \sqrt{3n_0\xi/2}$. For typical 1D BECs with $\xi \sim 0.7-1.0$ $\mu$m and $c_s\sim 1.0$ mm/s, we estimate $\tau_{\rm diffusion}\sim 0.05-0.1$ s, which reduces $\tau_{\rm qubit}$ in about 20\%.   \par
\begin{figure}[t!]
\includegraphics[scale=0.5]{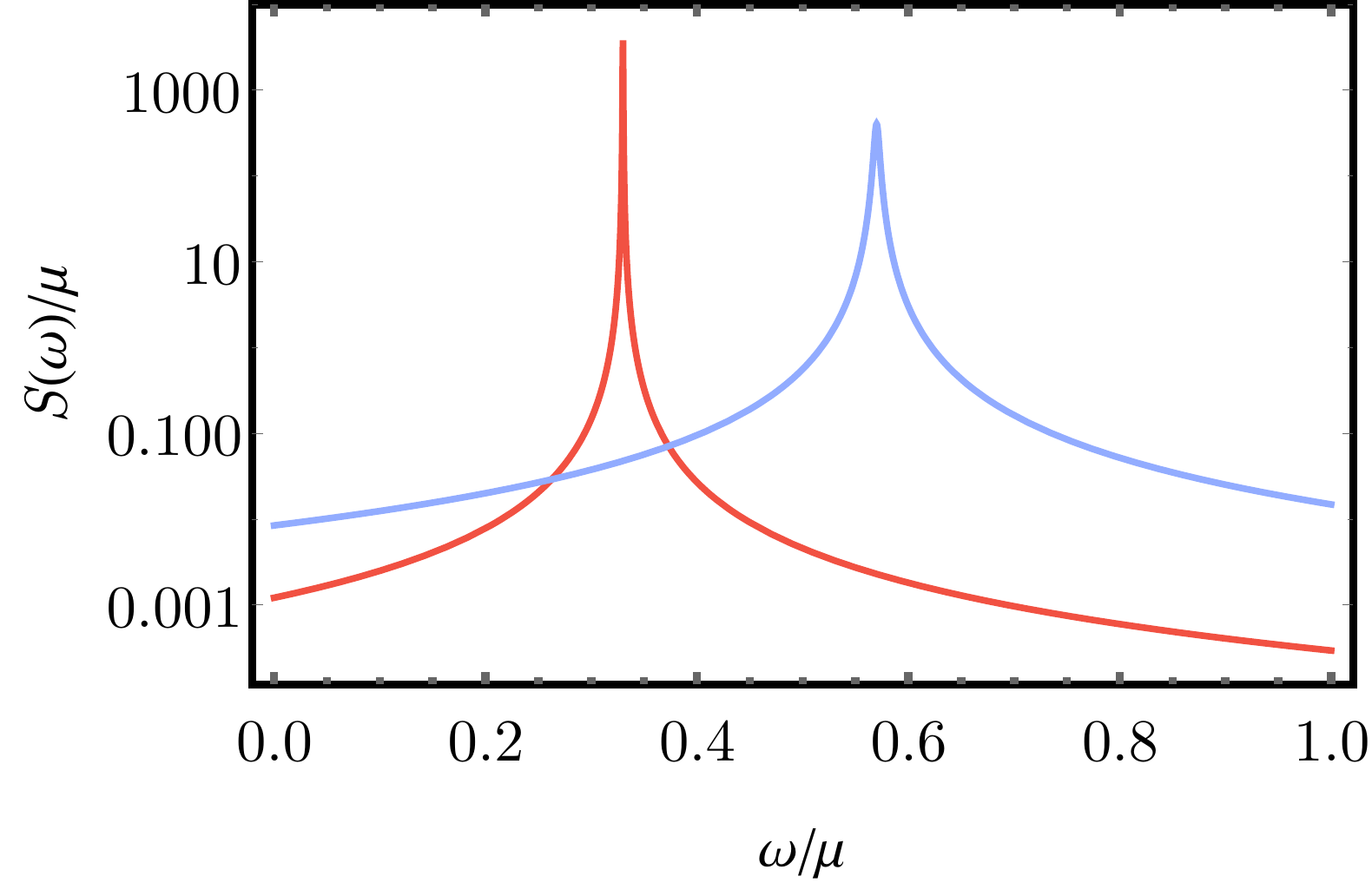}
\caption{(color online) Emission spectrum of a soliton qubit due to the interaction with the backround phonons. Red and blue curves are respectively obtained for $\nu =0.33$ and $\nu =0.79$.}
\label{fig_lorentzian}
\end{figure}

Finally, by putting Eqs. (\ref{coefficient b}) and (\ref{Coefficient a + gamma}) together, we can evalute the evolution of the amplitude coefficient $\beta_k(t)$  as
\begin{equation}
\beta_{k}(t) =\frac{i}{\hbar }g_{k}^{\ast }\int_{0}^{t}e^{-[\frac{\Gamma }{2}%
-i(\omega _{k}-\omega _{0})]t}dt,
\end{equation}
which yields the following Lorentzian spectrum
\begin{equation}
S(\omega_k)=\lim_{t\rightarrow \infty }\left\vert \beta_{k}(t)\right\vert ^{2} =\frac{1}{\hbar ^{2}}\frac{\left\vert g_{k}\right\vert ^{2}}{\frac{\Gamma ^{2}}{4}
+(\omega _{k}-\omega _{0})^{2}},  
\label{lorentzian}
\end{equation}
as illustrated in Fig. \ref{fig_lorentzian}. It is observed that the Lorentzian spectrum is narrower for a weak coupling constant $g_{12}$.
\section{Conclusion}
In conclusion, we have shown that a dark soliton in a quasi one-dimensional Bose-Einstein condensate can produce a well isolated two-level system, which can act as a matter-wave qubit of energy gap of a few kHz. This feature is intrinsic to the nonlinear nature of Bose-Einstein condensates and does not require manipulation of the internal degrees of freedom of the atoms. We observe that the decoherence induced by the quantum fluctuations (phonons) produce a finite qubit lifetime. Quite remarkably, leading calculations provide a qubit lifetime of the order of a few seconds, a time scale comparable to the duration of state-of-the art cold atomic traps. The only major limitation to the qubit robustness is the quantum diffusion of the soliton, which is estimated to reduce the qubit lifetime to around 20\% its value. This puts qubits made of dark solitons as good candidates to store information for large times ($\sim 0.01-1$ s), offering an appealing alternative to quantum optical of solid-state platforms. While dark solitons may not compete in terms of quantum scalability (the number of solitons in a typical elongated BEC is not expected to surpasse a few tens), their unprecedent coherence and lifetime will certainly make them attractive to the design of new quantum memories and quantum gates. Moreover, due to the possibility of interfacing cold atomic clouds with solid-state and optical systems, our findings may inspire further applications in hybdrid quantum computers.

\section*{Acknowledgements}

One of the authors (H. T.)  acknowledges the Security of Quantum Information Group for the hospitality and for providing the working conditions during the early stages of this work. Stimulating discussions with J. D. Rodrigues are acknowledged. The authors also thank the support from the DP-PMI programme and Funda\c{c}\~{a}o para a Ci\^{e}ncia e a Tecnologia (Portugal), namely through the scholarship number SFRH/PD/BD/113650/2015 and the grant number SFRH/BPD/110059/2015.E.V.C. acknowledges partial support from
FCT-Portugal through Grant No. UID/CTM/04540/2013.


\begin{thebibliography}{99}

\bibitem{henderson2009} K. Henderson, C. Ryu, C. MacCormick and M. G. Boshier, New
J. Phys. {\bf 11}, 043030 (2009).

\bibitem{donley2001} E. A. Donley, N. R. Claussen, S. L. Cornish, J. L. Roberts,
E. A. Cornell, C. E. Wieman, Nature {\bf 412}, 295 (2001).

\bibitem{leanhardt2003} A. E. Leanhardt, T. A. Pasquini, M. Saba, A. Schirotzek, Y. Shin, D. Kielpinski, D. E. Pritchard, W. Ketterle, Science {\bf 301}, 1513 (2003).

\bibitem{zoller98} D. Jaksch, C. Bruder, J. I. Cirac, C. W. Gardiner, P. Zoller, Phys. Rev. Lett. {\bf 81}, 3108 (1998).

\bibitem{greiner2002} M. Greiner, O. Mandel, T. Esslinger, T. W. Hänsch, I. Bloch, Nature {\bf 415}, 39 (2002).

\bibitem{orth2008} P. P. Orth, I. Stanic, and K. Le Hur, Phys. Rev. A {\bf 77}, 051601 (2008).

\bibitem{kalas2008} R. M. Kalas, A. V. Balatsky, D. Mozyrsky, Phys. Rev. B 	{\bf 78},184513 (2008).

\bibitem{santamore2008} D. H. Santamore, E. Timmermans, Phys. Rev. A {\bf 78}, 013619 (2008).

\bibitem {solenov2008a} D. Solenov and D. Mozyrsky, Phys. Rev. Lett. {\bf 100}, 150402 (2008).

\bibitem{solenov2008b}D. Solenov and D. Mozyrsky, Phys. Rev. A {\bf 78}, 053611 (2008)

\bibitem{kalas2010} R. M. Kalas, D. Solenov, E. Timmermans, Phys. Rev. A {\bf 81}, 053620 (2010).

\bibitem{porto2003} J. V. Porto, S. Rolston, B. Laburthe Tolra, C. J. Williams and W. D. Phillips, Phil. Trans. R. Soc. Lond. A {\bf 361}, 1417 (2003).

\bibitem{lundblad2009} N. Lundblad, J. M. Obrecht, I. B. Spielman, J. V. Porto, Nat. Phys. {\bf 5}, 575 (2009).

\bibitem{leggett89}  A. J. Leggett, {\it Quantum mechanics at the macroscopic level}, Ecole d'\'et\'e de physique th\'eorique (Les Houches, Haute-Savoie, France) (1986). 

\bibitem{dw_2008}
F. W. Strauch, M. Edwards, E. Tiesinga, C. Williams, and C. W. Clark Phys. Rev. A {\bf 77}, 050304(R) (2008).

\bibitem{andrews} M. R. Andrews, H.-J. Miesner, D. M. Stamper-Kurn, J. Stenger, and W. Ketterle, Phys. Rev. Lett. {\bf 82}, 2422 (1999).

\bibitem{tinkham2004}M. Tinkham, {\it Introduction to Supreconductivity}, Dover Publications, New York, 2nd Edn (2004).

\bibitem{solenov2011} D. Solenov and D. Mozyrsky, J. Comput. Theor. Nanosci. {\bf 8}, 481 (2011).

\bibitem{klein2007} A. Klein, M. Bruderer, S.R. Clark and D. Jaksch, New J. Phys. {\bf 9}, 411 (2007).

\bibitem{cirone2009} M. A. Cirone, G De Chiara, G. M. Palma and A. Recati, New J. Phys. {\bf 11}, 103055 (2009).

\bibitem{haikka2011} P. Haikka, S. McEndoo, G. De Chiara, G. M. Palma, and S. Maniscalco, Phys. Rev. A {\bf 84}, 031602 (2011).

\bibitem{mulansky2011} F. Mulansky, J. Mumford, and D. H. J. O’Dell, Phys. Rev. A {\bf 84}, 063602 (2011).

\bibitem{peotta2013} S. Peotta, D. Rossini, M. Polini, F. Minardi, R. Fazio, Phys. Rev. Lett. {\bf 110}, 015302 (2013).

\bibitem{stanigel2012} K. Stannigel, P. Rabl, and P. Zoller, New J. Phys. {\bf 14}, 063014 (2012).

\bibitem{ramos2014} T. Ramos, H. Pichler, A. J. Daley, P. Zoller, Phys. Rev. Let. {\bf 113}, 237203 (2014). 

\bibitem{petersen2014} J. Petersen, J. Volz, and A. Rauschenbeutel, Science {\bf 346}, 67 (2014).

\bibitem{mitsch2014} R. Mitsch, C. Sayrin, B. Albrecht, P. Schneeweiss, and A. Rauschenbeutel, Nature Communication \textbf{5}, 5713 (2014).

\bibitem{sollner2014} I. Söllner, S. Mahmoodian, A. Javadi, and P. Lodahl,  Nature Nanotechnology \textbf{10}, 775 (2015).

\bibitem{young2014} A. B. Young, A. Thijssen, D. M. Beggs, L. Kuipers, J. Rarity, and R. Oulton,  Phys. Rev. Lett. \textbf{115}, 153901 (2015).

\bibitem{ramos2016} T. Ramos, B. Vermersch, P. Hauke, H. Pichler,  and P. Zoller, Phys. Rev. A {\bf 93}, 062104 (2016).

\bibitem{vermersch2016} B. Vermersch, T. Ramos, P. Hauke, and P. Zoller, Phys. Rev. A {\bf 93}, 063830 (2016). 

\bibitem{luiz} F. S. Luiz, E. I. Duzzioni, L. Sanz, Brazilian Journal of
Physics \textbf{45}, 550 (2015).

\bibitem{kivshar} Y. S. Kivshar and G. P. Agrawal, \textit{Optical Solitons:
From Fibers to Photonic Crystals} (Academic Press, San Diego, USA, 2003).

\bibitem{burger} S. Burger, S. Dettmer, W. Ertmer, K. Sengstock, A. Sanpera,
G. V. Shlyapnikov, and M. Lewenstein, Phys. Rev. Lett. \textbf{83}, 5198
(1999).

\bibitem{Denschlag} J. Denschlag, J. E. Simsarian, D. L. Feder, C. W. Clark,
L. A. Collins, J. Cubizolles, L. Deng, E. W. Hagley, K. Helmerson, W. P.
Reinhart, S. L. Rolston, B. I. Schneider, and W. D. Phillips, Science 
\textbf{287}, 97 (2000).

\bibitem{anderson} B. P. Anderson, P. C. Haljan, C. A. Regal, D. L. Feder,
L. A. Collins, C. W. Clark, and E. A. Cornell, Phys. Rev. Lett. \textbf{86},
2926 (2001).

\bibitem{Krakel} D. Krakel, N. J. Halas, G. Giuliani, and D. Grischkowsky,
Phys. Rev. Lett. \textbf{60}, 29 (1988); G. A. Swartzlander, D. R. Andersen,
J. J. Regan, H. Yin, and A. E. Kaplan, ibid. \textbf{66}, 1583 (1991).

\bibitem{Denardo} B. Denardo, W. Wright, S. Putterman, and A. Larraza, Phys.
Rev. Lett. \textbf{64}, 1518 (1990).

\bibitem{Chen} M. Chen, M. A. Tsankov, J. M. Nash, and C. E. Patton, Phys.
Rev. Lett. \textbf{70}, 1707 (1993).


\bibitem{Dutton} Z. Dutton, M. Budde, C. Slowe, and L.V. Hau, Science 
\textbf{293}, 663 (2001).

\bibitem{Reinhardt} W. P. Reinhardt and C. W. Clark, J. Phys. B \textbf{30},
L785 (1997).

\bibitem{Scott} T. F. Scott, R. J. Ballagh, and K. Burnett, J. Phys. B 
\textbf{31}, L 329 (1998)

\bibitem{jackson} B. Jackson, N. P. Proukakis, C. F. Barenghi, Phys. Rev. A 
\textbf{75}, 051601 (2007).

\bibitem{dziarmaga} J. Dziarmaga, Z. P. Karkuszewski, and K. Sacha, J. Phys.
B: At. Mol. Opt. Phys. \textbf{36}, 1217 (2003).

\bibitem{gael2005} G. A. El and A. M. Kamchatnov, Phys. Rev. Lett. {\bf 95},
204101 (2005).

\bibitem{tercas} H. Ter\c{c}as, D. D. Solnyshkov and G. Malpuech, Phys. Rev.
Lett. \textbf{110}, 035302 (2013); ibid \textbf{113}, 036403 (2014).

\bibitem{perez} V. M. Perez-Garcia, H. Michinel and H. Herrero, Phys. Rev. A 
\textbf{57}, 3837 (1998).

\bibitem{carr} L. D. Carr, C. W. Clark and W. P. Reinhardt, Phys. Rev. A 
\textbf{62}, 063611 (2000).

\bibitem{Muryshev} A. Muryshev, G. V. Shylapnikov, W. Ertmer, K. Sengstock,
and M. Lewenstein, Phys. Rev. Lett. \textbf{89}, 110401 (2002).

\bibitem{huang} G. Huang, J. Szeftel, and S. Zhu, Phys. Rev. A \textbf{65},
053605 (2002).

\bibitem{zakharov72} V. E. Zakharov and A. B. Shabat, Sov. Phys. JETP 
\textbf{34}, 62 (1972).

\bibitem{zakharov73} V. E. Zakharov and A. B. Shabat, Sov. Phys. JETP 
\textbf{37}, 823 (1973).

\bibitem{parker} N. Parker, Numerical Studies of Vortices and Dark Solitons in atomic Bose Einstein Condensates, Ph.D Thesis (2004).

\bibitem{pelinovsky} D. E. Pelinovsky, Y. S. Kivshar, and V. V. Afanasjev, Phys. Rev. E \textbf{54}, 2015 (1996).

\bibitem{wadkin} D. C. Wadkin-Snaith and D. M. Gangardt, Phys. Rev. Lett. \textbf{108}, 085301 (2012).

\bibitem{john} J. Leknera, Am. J. Phys. \textbf{75}, \textbf{12} (2007).

\bibitem{Dziarmaga04} J. Dziarmaga, Phys. Rev. A \textbf{70}, 063616 (2004).

\bibitem{pitaevskii} L. Pitaevskii and S. Stringari, \textit{Bose-Einstein
Condensation} (Clarendon, Oxford, 2003).

\bibitem{pethick} C. J. Pethick and H. Smith, \textit{Bose Einstein
Condensation in Dilute Gases}; Second Edition (Cambridge University Press,
Cambridge, England, 2008).
















\end{thebibliography}
\end{document}